\documentclass[aip,preprint,a4paper]{revtex4-1} 
\usepackage{amssymb}
\usepackage{amsmath}
\usepackage{color}
\usepackage{hyperref}
\usepackage{bm}
\usepackage{epsfig}
\usepackage{subcaption}
\usepackage{float}

\begin{document}
\title{Simulation of one and two qubit superconducting 
quantum gates under the non-Markovian $1/f$ noise}
\author{Yinjia Chen}
\author{Shuocang Zhang}
\author{Qiang Shi}\email{qshi@iccas.ac.cn}
\affiliation{Beijing National Laboratory for
Molecular Sciences, State Key Laboratory for Structural Chemistry of
Unstable and Stable Species, Institute of Chemistry, Chinese Academy
of Sciences, Zhongguancun, Beijing 100190, China, and
University of Chinese Academy of Sciences, Beijing 100049, China}

\begin{abstract}

Non-Markovian $1/f$ noise consists a dominant source of
decoherence in superconducting qubits, yet its slow nature
poses a significant challenge for accurate simulation.
Here we develop a hierarchical equations of motion (HEOM) framework that enables efficient and reliable modeling of qubit dynamics and gate operations under $1/f$ noise. 
By using the approach, it is first shown that perturbative quantum master equations may fail to reproduce the correct dephasing dynamics of a qubit coupled to slow baths. We then analyze dynamical decoupling sequences by including effects 
of finite pulse duration. It is found that different 
pulse sequences results in different behavior in error accumulation: all X-CPMG sequences exhibit linear scaling with parity effects, Y-CPMG follows quadratic growth, and alternating XY-type sequences can suppress the error accumulation significantly. Finally, we extend the framework to two-qubit cross-resonance (CR) gates, reconstructing the full Choi matrix and Pauli Transfer Matrix (PTM) to identify the incoherent error induced by $1/f$ noise. Together, these results establish HEOM as a robust methodology for simulating the environmental noise in superconducting circuits and provide new insights into error mechanisms in both single- and two-qubit gates.

\end{abstract}

\maketitle

\section{Introduction}

Over the past two decades, quantum computers based on superconducting qubits have made remarkable progress, achieving steadily increasing coherence times, high-fidelity gate operations, and the integration of tens to over one hundred qubits on a single chip \cite{acharya25,gao25}. Nevertheless, in the noisy intermediate-scale quantum (NISQ) era, overcoming decoherence from environmental noise remains a central obstacle to realizing fault-tolerant quantum computation \cite{preskill18}. It is therefore important to establish reliable methods for simulating 
quantum-gate operations in realistic noisy environments and to analyze the interplay between external driving fields and environmental dissipation.

To model dissipative effects in qubits and quantum circuits, many theoretical studies have relied on simplified error models based on Lindblad master equations \cite{lidar13}. These approaches assume Markovian dynamics, which neglect temporal correlations and therefore cannot fully capture the complexity of real device environments. In superconducting qubits in particular, a growing body of experimental and numerical evidence points to the importance of non-Markovian effects. Examples include 
oscillatory qubit purity decay \cite{quintana17,burkard23,greenaway24} and gate-dependent
error accumulation \cite{proctor20}, both of which are signatures of noise with temporal memory \cite{agarwal24}. Together, these findings demonstrate that non-Markovian noise plays a critical role in the decoherence process and highlight the need for simulation frameworks that can go beyond the Markovian approximation.

Among the various noise mechanisms affecting superconducting qubits, $1/f$ noise \cite{siddiqi21} poses a particularly severe challenge. It is routinely observed in superconducting circuits using quantum noise spectroscopy (QNS) techniques \cite{bylander11}, where the measured power spectral density (PSD) in the low frequency range typically follows $S(\omega) \propto 1/\omega^{\alpha}$ with $\alpha \approx 1$. 
In the literature, the origin of $1/f$ noise has been associated with surface surface spins \cite{laforest15,rower23} in the case of flux noise, two-level fluctuators (TLFs) arising from material defects \cite{muller19}, and quasiparticle dynamics \cite{liu24a,vepsalainen20}. Its predominance at low frequencies causes slow fluctuations in qubit parameters, resulting in dephasing and degraded gate fidelities. Moreover, this type of
noise is currently not effectively suppressed by
conventional error correction strategies
\cite{lidar13,devitt13}, which are primarily designed to
handle high-frequency noise.

It is therefore essential to develop theoretical methods capable of treating low-frequency noise sources within a fully non-Markovian framework. Quantum master equations (QMEs) \cite{breuer07a,burkard23,wang23,burkard09,gulcsi24} can in principle capture non-Markovian effects by perturbatively treatment of the system–bath coupling. However, their accuracy quickly deteriorates for slow baths with long correlation times, even in the simplest case of single-qubit pure dephasing (see Sec.~\ref{sec:single_qubit_results} for an example). On the other hand, the hierarchical equations of motion (HEOM) method \cite{tanimura20} provides a numerically exact description of open quantum system dynamics. HEOM has already been applied to a range of quantum computing problems, including the calculation of decoherence times $T_1/T_\phi$ \cite{nakamura24}, modeling entanglement dynamics \cite{dijkstra10,ma12}, simulating quantum algorithms \cite{zhang24}, and analyzing bang-bang pulse control \cite{nakamura24a,nakamura25}.

Recent works by \citeauthor{nakamura24a} \cite{nakamura24a,nakamura25} established an important HEOM-based framework for systematically studying the impact of non-Markovian noise on single-qubit dynamics. Different noise environments including Ohmic and sub-Ohmic with various power-law exponents are compared.  Ref.~\cite{nakamura24a} also rigorously examined decoherence suppression using Carr-Purcell-Meiboom-Gill (CPMG) sequences, although the analysis was restricted to idealized, instantaneous pulses. Moreover, both studies focused exclusively on single-qubit systems, leaving open the question of how non-Markovian noise affects two-qubit entangling gates.

In this work, we extend HEOM-based simulations of qubit operations to bridge these gaps. First, we focus on the $1/f$ noise model, which is more representative of experimental observations, and more challenging to handle numerically. We introduce a strategy to treat the associated computational difficulties by introducing a low-frequency cutoff and treat the ultra slow component of the environmental noise by static disorder. Second, we investigate the interplay between $1/f$ noise and realistic, finite-duration control pulses in single-qubit CPMG sequences, analyzing how error accumulation differs for different types of gate operations. Third, we broaden the scope to two-qubit cross-resonance (CR) gates by developing a full simulation pipeline that includes Choi matrix reconstruction and Pauli Transfer Matrix (PTM) analysis, enabling a detailed characterization of the error patterns induced by non-Markovian noise.

The remainder of this article is organized as follows.
Section~\ref{sec:Methods} introduces the theoretical framework, including the open-system models for dissipative qubits and the HEOM treatment of $1/f$ noise.
Section~\ref{sec:Results} presents the main findings: we first benchmark the HEOM framework against analytical solutions for single-qubit pure dephasing, and show the failure of perturbative methods to capture effects of the slow bath. The error patterns of CPMG sequences using different combinations of finite-duration X- and Y-gates are then investigated. Finally, we simulate the CR gate under $1/f$ noise and quantify its performance through Choi matrix and PTM analysis. Conclusions and discussions are made in Section~\ref{sec:conclusion}.

\section{Methods}\label{sec:Methods} 
\subsection{Model Hamiltonian}
\label{sec:modelhamiltonian}

We first introduce the model Hamiltonian for a single
dissipative qubit. A superconducting qubit is a complex
quantum system subject to multiple noise mechanisms with
distinct spectral
signatures\cite{paladino19,krantz19,muller19}. The
population relaxation time $T_1$ is primarily determined by
the power spectral density at the qubit transition
frequency\cite{schriefl06} (typically in the GHz range). By
contrast, $1/f$ noise is dominated by low-frequency
components (MHz and below) and mainly contributes to
dephasing rather than relaxation\cite{paladino19,krantz19}.

Currently, there are several different approaches to
describing the $1/f$
noise\cite{schriefl06,tripathi24,faoro15,shnirman05}.  
In the quasi-static approximation \cite{tripathi24}, the
low-frequency PSD is taken as approximately constant over
short timescales, and the static noise is modeled as a
Gaussian random variable. An alternative is to represent
$1/f$ noise as an ensemble of two-level fluctuators (TLFs)
\cite{schriefl06,faoro15,shnirman05,Ithier05}, with
appropriately broad parameter distributions that yield an
aggregate $1/f$ spectrum. In this work, we adopt a
harmonic-bath (spin-boson) description\cite{clerk10} with a
spectral density tailored to reproduce the $1/f$ PSD. This
linear-bath framework is standard in open quantum system
theory\cite{leggett87,weiss12} and offers practical
advantages to include temperature and non-Markovian effects.

Within the harmonic bath framework,\cite{leggett87,
weiss12,clerk10} the total Hamiltonian is divided into three
parts: $H_S$ for the qubit, $H_B$ for the bath, and $H_{SB}$
for the system-bath coupling:
\begin{equation}
H = H_S + H_B + H_{SB} \;\;.
\end{equation}
The specific form of the system Hamiltonian, $H_S$, which includes free evolution and external drives, will be defined for the single- and two-qubit cases in the relevant sections. We assume that each qubit is coupled independently to its own local bath, leading to the following general forms for the bath and interaction Hamiltonians:
\begin{equation}\label{equ:hbath_general}
  H_B = \sum_i \sum_j \left[\frac{\left(p_{j}^{(i)}\right)^2}{2m_{j}^{(i)}} 
  + \frac{1}{2}m_{j}^{(i)}\left(\omega_{j}^{(i)}\right)^2 \left(x_{j}^{(i)} 
  - \frac{c_{j}^{(i)}}{m_{j}^{(i)}\left(\omega_{j}^{(i)}\right)^2} \sigma_z^{(i)}\right)^2\right] \;\;,
\end{equation}
\begin{equation}\label{equ:hsb_general}
  H_{SB} = \sum_i \left( \sum_j c_{j}^{(i)}x_{j}^{(i)} \right) \otimes \sigma_z^{(i)} \;\;.
\end{equation}
Here, the index $i$ runs over the qubits in the system, and $j$ indexes the harmonic oscillator modes of the bath coupled to the $i$-th qubit. $p_{j}^{(i)}, \omega_{j}^{(i)}, x_{j}^{(i)}$, and $m_{j}^{(i)}$ denote the momentum,
frequency, position, and mass of the $j$-th
harmonic-oscillator mode of the $i$-th bath. $c_{j}^{(i)}$ is the corresponding coupling constant, and $\sigma_z^{(i)}$
is the Pauli Z matrix acting on the $i$-th qubit. Throughout this paper, we set
$\hbar = 1$.

Since the main focus of this work is to present an efficient
approach to study the impact of slow $1/f$ noise, we use
exclusively the $Z$-type noise in the current study. The
system-bath interaction for the $i$-th qubit is fully characterized by its
spectral density $J^{(i)}(\omega)$, defined as:
\begin{equation}
  \label{eq:jomega}
  J^{(i)}(\omega) = \frac{\pi}{2}\sum_{j} \frac{\left(c_j^{(i)}\right)^2}{m_{j}^{(i)} \omega_{j}^{(i)}}
  \delta\left(\omega- \omega_{j}^{(i)}\right) \;\;.
\end{equation}
The corresponding PSD of the bath operator fluctuations for the $i$-th bath is related to the spectral
density via the fluctuation-dissipation theorem:
\begin{equation}
\label{eq:somega}
  S^{(i)}(\omega) = \frac{J^{(i)}(\omega)}{1-e^{-\beta\omega}} \;\;.    
\end{equation}

For simplicity, in this work we assume that all qubits are subject to identical, independent noise sources. This allows us to drop the superscript $(i)$, such that $J^{(i)}(\omega) = J(\omega)$ for all qubits. We consider a spectral density of the 
following form\cite{nakamura24,nakamura24a,nakamura25}:
\begin{equation}
\label{eq:1overf}
J(\omega) = \frac{\pi}{2}\text{sgn}(\omega)
\frac{\eta \omega_q^{1-s}|\omega|^s}{(1+(\omega/\omega_{hc})^2)^2},
\end{equation}
where $s$ is the spectral exponent, with $s=0$ corresponding to $1/f$
noise. $\eta$ denotes the dimensionless system-bath coupling strength, and
$\omega_{hc}$ is the high-frequency cutoff. $\omega_q$ is a
characteristic frequency of the system, which we set to the qubit frequency $\omega_1=5$ GHz.
 
The sign function in Eq.~\eqref{eq:1overf} extends the definition 
of the spectral density to $\omega <0$ by enforcing $J(\omega)=-J(-\omega)$. 
This leads to the following expression for the bath correlation function
$C(t)\equiv \langle F^{(i)}(t) F^{(i)}(0) \rangle$, where $F^{(i)} = \sum_j c_j^{(i)} x_j^{(i)}$ is the collective bath operator for the $i$-th qubit. Due to the identical bath assumption, this correlation function is the same for all qubits:
\begin{equation}
  \label{eq:ct}
  C(t) = \frac{1}{\pi}\int_{-\infty}^{\infty}\!\!d\omega e^{-i\omega t}
 S(\omega)=\frac{1}{\pi}\int_{-\infty}^{\infty}\!\!d\omega
  \frac{e^{-i\omega t}J(\omega)}{1-e^{-\beta\hbar\omega}}\;.
\end{equation}

When $s=0$, the spectral density in Eq.~\eqref{eq:1overf}
gives the correct $1/f$ behavior at low frequencies.
However, there is a technical difficulty in using this
spectral density in numerical simulations, as
$S(\omega)$ diverges at $\omega=0$. To solve this problem,
we introduce a low-frequency cutoff $\omega_{lc}$, and the
spectral density is modified as:
\begin{equation}
\label{eq:1overf2}
  J(\omega) = \frac{\pi}{2}\text{sgn}(\omega)
  \frac{\eta \omega_q^{1-s}|\omega|^s}{(1+(\omega/\omega_{hc})^2)^2}
  \left[\theta(\omega-\omega_{lc})+\theta(-\omega-\omega_{lc})\right],
\end{equation}
where $\theta(x)$ is the soft-Heaviside step function, which is 
given by:
\begin{equation}\label{eq:cutoff}
    \theta(x)=1-\frac{1}{1+e^{\frac{x}{\Phi}}}
\end{equation}
with $\Phi = \omega_{lc}/10$ representing the transition
width of the step function. Here, $\Phi$ controls the
``sharpness" of $\omega_{lc}$ to improve efficiency in
decomposing the bath correlation function $C(t)$ (see
Sec~\ref{sec:HEOM}).

The low-frequency cutoff $\omega_{\text{lc}}$ can be chosen
to be inversely proportional to the measurement time $t_m$,
that is, $\omega_{\text{lc}} \sim 1/t_m$ \cite{niemann13}.
This implies that the spectral density with low-frequency
cutoff in Eq.~\eqref{eq:1overf2}
neglects noise components with
fluctuation timescales longer than $t_m$. In practical
simulations, we set $t_m$ to be between $1$ and $10^6$
times the qubit oscillation period $1/\omega_1$, which
corresponds to $\omega_{\text{lc}}/2\pi$ in the range of 1
Hz to 1 MHz. To account for
extremely low-frequency noise, i.e., the difference 
between the spectral density with (Eq.\eqref{eq:1overf2}) and
without (Eq.\eqref{eq:1overf}) the low-frequency cutoff,
we treat their difference as Gaussian static disorder.

\subsection{The HEOM method}
\label{sec:HEOM}

The HEOM is a powerful and widely used tool for simulating
open quantum system dynamics \cite{tanimura89,tanimura20}.
It proceeds by expressing the bath correlation function
$C(t)$ as a sum of exponentials \cite{dan23a}:
\begin{align}
\label{eq:ct2}
C(t) = \frac{1}{2\pi}\int_{-\infty}^{\infty}
S(\omega)e^{-i\omega t}d\omega = \sum_k d_k e^{-\gamma_k t} \;\;,
\end{align}
after which the corresponding HEOM (for
a single qubit) can be derived as \cite{xu22}:  
\begin{align}\label{equ:extendedHEOM}
  \begin{split}
    \frac{d\hat{\tilde{\rho}}_{\mathbf{m,n}}}{dt} = & \left(-i\mathcal{L} + 
    \sum_k m_k \gamma_k + \sum_k n_k \gamma_l^*\right) 
    \hat{\tilde{\rho}}_{\mathbf{m,n}} \\
    & -i \sum_k \sqrt{(m_k + 1) d_k}[\hat{q}, \hat{\tilde{\rho}}_{\mathbf{m}_k^+, \mathbf{n}}] 
    - i \sum_k \sqrt{(n_k + 1) d_k^*}[\hat{q}, \hat{\tilde{\rho}}_{\mathbf{m}, \mathbf{n}_k^+}] \\
    & -i \sum_k \sqrt{m_k d_k} \hat{q} \hat{\tilde{\rho}}_{\mathbf{m}_k^-, \mathbf{n}} 
    + i \sum_k \sqrt{n_k d_k^*} \hat{\tilde{\rho}}_{\mathbf{m}, \mathbf{n}_k^-} \hat{q} \;\;.
  \end{split}
\end{align}

Here, $\gamma_k$s are assumed to be complex,
and the system operator $\hat{q} = \sigma_z$.
The multi-index vectors $\mathbf{m,n}$ label different
tiers in the hierarchy, with their components corresponding
to different exponential expansion terms. The notation
$\mathbf{m}_k^+$ ($\mathbf{m}_k^-$) represents the vector
obtained from $\mathbf{m}$ by increasing (decreasing) its
k-th component by one while keeping other components
unchanged, and similarly for $\mathbf{n}_k^+$ and
$\mathbf{n}_k^-$. The Liouville superoperator $\mathcal{L}$
defines the free evolution of the system through
$\mathcal{L}\hat{\rho} = [H_S, \hat{\rho}]$

An important challenge in applying HEOM is the efficient
decomposition of the bath correlation function $C(t)$. This
is especially critical for $1/f$ noise in superconducting
circuits: the spectrum diverges toward zero frequency and is
not well captured by standard Padé \cite{hu11} or Matsubara
\cite{tanimura20} schemes. Moreover, because superconducting
qubits typically operate at low temperature and because of
the cutoff function in Eq.\eqref{eq:cutoff}, the expansion
in Eq.(\ref{eq:ct2}) may require a large number of
exponential terms to converge. Consequently, a compact and
accurate decomposition is crucial for efficient HEOM
simulations.

To address this problem, we employ the frequency-domain
barycentric spectral decomposition (BSD) method
\cite{xu22,dan23a}, which has proven to be very 
effective for low temperature simulations. 
Other efficient schemes for
decomposing $C(t)$ have also been proposed recently
\cite{takahashi24}. To further reduce the computational cost
when many exponential terms are involved, we use the
Matrix Product State (MPS) method combined with the
time-dependent variational principle (TDVP) algorithm to
propagate the HEOM. More details of the MPS-TDVP 
methods can be found in Refs.~\cite{shi18,guan24}.
 

\subsection{Equation of motion for static disorder}

This paper investigates the impact of static disorder on qubit dynamics using two complementary modeling approaches.
The first is the cutoff plus static disorder scheme, described in Sec. II A. In this scheme, only the sub-cutoff portion of the noise spectrum, with variance $\sigma^2 = \int_0^{\omega_{lc}} S(\omega) d\omega$, is treated as static noise. This contribution is incorporated into the HEOM formalism as a zero-frequency mode \cite{gelin21,huang24}, so that the bath correlation function is expressed as a sum of a dynamic part and a constant term: $C(t) \rightarrow C_{\omega > \omega_{lc}}(t) + \sigma^2$. Within the HEOM formalism in  Eq.\eqref{equ:extendedHEOM}, the constant term $\sigma^2$ is represented as an additional exponential with zero decay rate $\gamma_0 = 0$ and amplitude $d_0 = \sigma^2$. For simplicity, detailed form of the modified HEOM will not be presented. This hybrid model is designed to provide a controlled way of studying how different choice of the low-frequency cutoff $\omega_{lc}$ affects qubit decoherence dynamics.

The second approach is the total static disorder scheme, which serves as a simplified benchmark by approximating the entire noise spectrum as a single Gaussian-distributed static disorder. In this model, the bath correlation function reduces to a constant equal to the total noise power, $C(t) = \langle \delta\omega_z^2 \rangle$. Correspondingly, the HEOM formalism reduces to a much simpler structure. For a single static mode, it takes the form:
\begin{align}\label{equ:HEOM_static_disorder}
    \frac{d\hat{\tilde{\rho}}_{n}}{dt} =  -i\mathcal{L}
    \hat{\tilde{\rho}}_{n}- i \sqrt{\langle\delta\omega_z^2\rangle} \left( [\hat{q}, \hat{\tilde{\rho}}_{n+1}] + [\hat{q}, \hat{\tilde{\rho}}_{n-1}] \right) 
   \;\;,
\end{align}
where $n$ denotes the hierarchy index for the static mode and $\hat{q} = \sigma_z$.
Unless stated otherwise in Sec.\ref{sec:single_qubit_results}, all subsequent references to “static disorder” in this paper—including the analysis of dynamical decoupling (Sec.\ref{sec:DD}) and the CR gate (Sec.\ref{sec:CR})—refer to this total static disorder model, simulated using the simplified equation of Eq.\eqref{equ:HEOM_static_disorder}.

The above approach, which incorporates static disorder as an additional degree of freedom within the HEOM hierarchy, eliminates the need to perform ensemble averaging over many independent simulations, thereby reducing statistical uncertainty. It also introduces only minimal computational overhead compared to standard HEOM calculations, since only one single zero-frequency mode is added.

\section{Results}
\label{sec:Results}

\subsection{Dephasing dynamics of a single qubit}
\label{sec:single_qubit_results}

We begin by modeling a single superconducting qubit, with system parameters typical of state-of-the-art fixed frequency transmons. The qubit frequency is set to $\omega_1/2\pi = 5~\mathrm{GHz}$ \cite{bylander11}, and 
$H_S = \frac{\omega_1}{2}\sigma_z$.
The qubit is coupled to a bath that gives rise to $1/f$ noise, characterized by the parameters in Table~\ref{tab:bath_params}. We set the high-frequency cutoff $\omega_{hc}/2\pi = 10~\mathrm{GHz}$, the low-frequency cutoff $\omega_{lc}$ is to be specified, and the dimensionless system-bath coupling strength is $\eta = 10^{-7}$. With this spectral density, the dephasing time $T_\phi$ can be estimated by assuming an equivalent Gaussian static disorder \cite{Ithier05,Astafiev04}:
\begin{equation}
\left(\frac{1}{T_\phi}\right)^2 = 
\frac{\pi}{2}\int_{-\infty}^{\infty} S(\omega)  d\omega \;\;,
\end{equation}
which yields $T_\phi \approx 200~\mathrm{ns}$. The temperature is $T = 50~\mathrm{mK}$.


The simplest exactly solvable framework for quantum
decoherence is the single-qubit pure dephasing model
\cite{breuer07a,skinner82}, where one starts from the 
$|+\rangle$ state of the qubit and calculates the 
decay of the off-diagonal element of the density matrix
$\rho_{01}(t) = \langle 0|\rho(t)|1\rangle$.  
This model is first used to analyze the effect of static disorder and help us to choose the low-frequency cutoff
$\omega_{lc}$ for balanced accuracy and efficiency. 
In the pure dephasing model, the decay of $\rho_{01}(t)$ can be calculated analytically as\cite{breuer07a,skinner82,skinner86b,Ithier05,Astafiev04}: \begin{equation}\label{eq:pure_dephasing_decoherence}
  \rho_{01}(t)=\rho_{01}(0)e^{-i\omega t}e^{-\Gamma(t)} \;\;,
\end{equation}
where the time-dependent $\Gamma(t)$ is determined by the  the spectral density $J(\omega)$:
\begin{equation}
  \Gamma(t) = \frac{4}{\pi}\int_0^\infty d\omega~ 
  \frac{J(\omega)}{{\omega^2} }\text{coth}\left(\frac{\beta\omega}{2}\right)
  (1-\cos{\omega t}) \;\;.
\end{equation}

\begin{figure}[H]
    \centering
    \includegraphics[width=\linewidth]{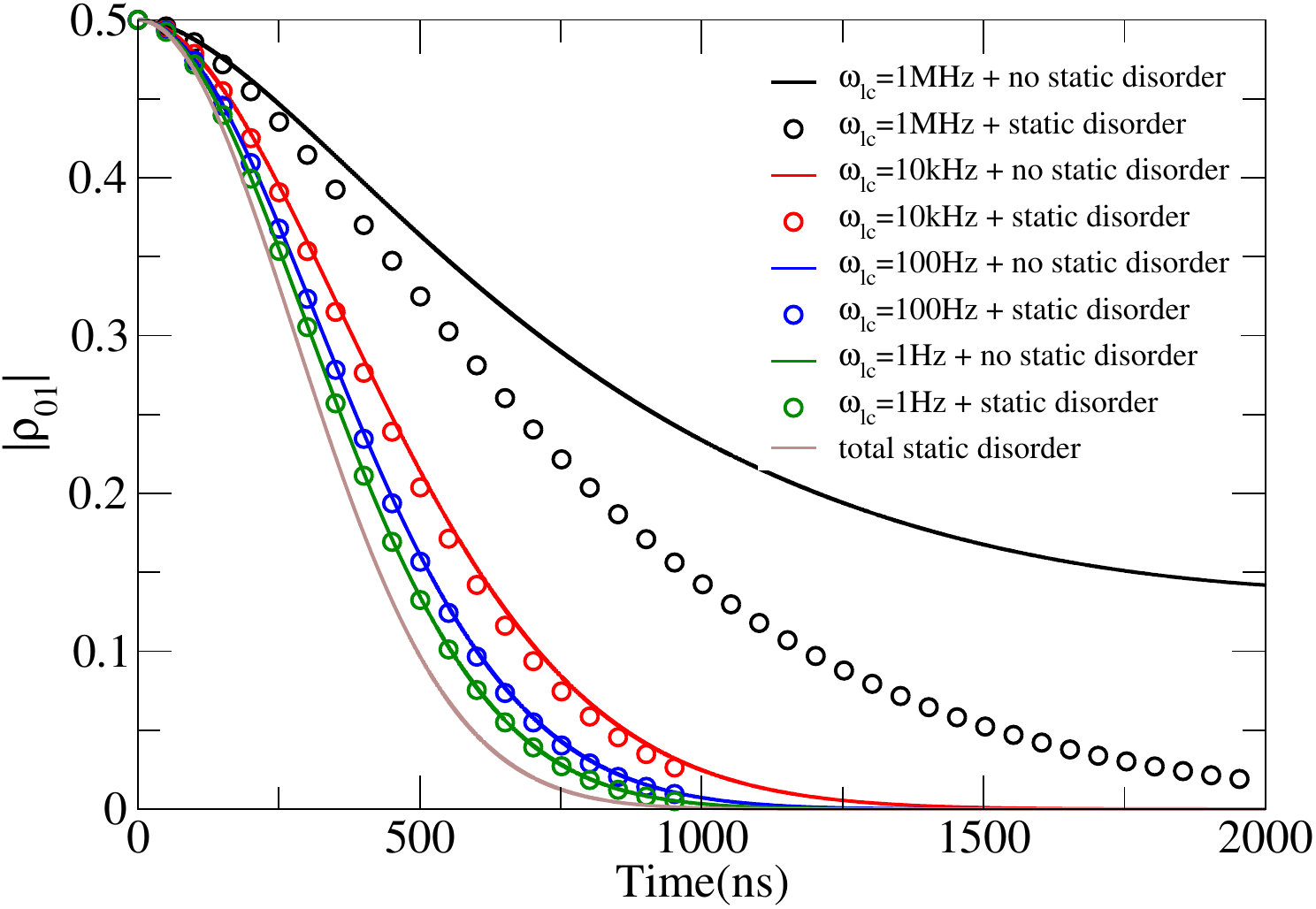}
    \caption{Dynamics of the coherence of a single qubit ($|\rho_{01}(t)|$) 
    initially in the state $|+\rangle$ under different
    $\omega_{lc}$ and PSD definitions. The red solid line
    represents the coherence decay trajectory from the black
    line in Fig~\ref{fig:compare_qme}}
    \label{fig:static_disorder}
\end{figure}

In Fig.~\ref{fig:static_disorder}, we use this analytical solution to study the pure dephasing dynamics for several representative cutoff values: $\omega_{lc}/2\pi = 1~\mathrm{MHz}, 10~\mathrm{kHz}, 100~\mathrm{Hz},$ and $1~\mathrm{Hz}$. For each cutoff, we compare two schemes to treat the PSD below $\omega_{lc}$: a cutoff only scheme where noise components below $\omega_{lc}$ are simply neglected (solid lines), and a cutoff plus static disorder scheme where the below cutoff components are treated as a static Gaussian noise with $S(\omega)=S(\omega_{lc})$ (open circles). For a fair comparison, the coupling strength $\eta$ is adjusted for each cutoff to maintain a constant total integrated noise power, with the specific parameter values listed in Table~\ref{tab:cutoff_params}. The results are also compared with a total static disorder model, where the entire noise spectrum is approximated as static (brown line).

As shown in Fig.~\ref{fig:static_disorder}, the decoherence dynamics are sensitive to the choice of the low-frequency cutoff $\omega_{lc}$ and the treatment of below cutoff noise. When the cutoff is high ($\omega_{lc}/2\pi = 1$ MHz), there is a significant discrepancy between the cutoff only (black solid line) and the cutoff plus static disorder (black circles) models, indicating that neglecting the below cutoff components is a poor approximation. As $\omega_{lc}$ is lowered into the kHz range and below, the dynamics of the two schemes become nearly indistinguishable (see, e.g., the red, blue, and green solid lines and open circles). 
It is also observed that, as $\omega_{lc}$ decreases, the coherence decay curve approaches more closely towards the result predicted by the total static disorder model, in consistent with the low-frequency nature of $1/f$ noise.

Based on the above analysis, we need to choose a  $\omega_{lc}$ that is both physically sound and numerically practical for the HEOM simulations (Eq.~\eqref{equ:extendedHEOM}). We find that when $\omega_{lc}$ is in the kHz range, the estimated dephasing time $T_\phi$ remains stable within the same order of magnitude. In addition, choosing an excessively small cutoff (e.g., in the Hz range) can lead to numerical instabilities in for efficient HEOM simulations with the BSD method. Therefore, we select $\omega_{lc}/2\pi = 10~\mathrm{kHz}$ for all subsequent simulations, as it provides balance between accuracy and numerical stability. With this choice of $\omega_{lc}$, the parameters of the dissipative qubit is given in Table~\ref{tab:bath_params}.

\begin{figure}[H]
\centering
\includegraphics[width=\linewidth]{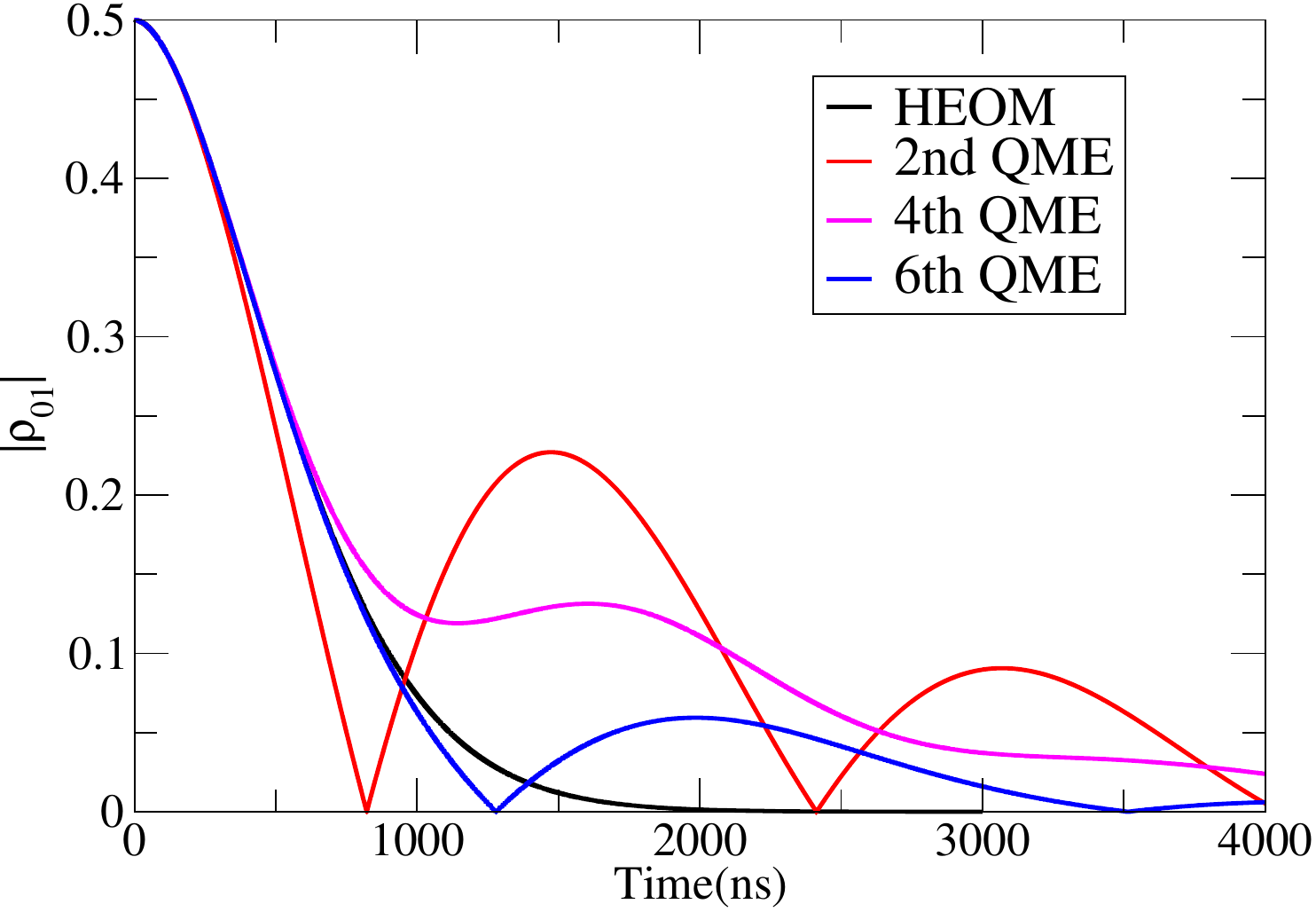}
\caption{Dynamics of the coherence of a single qubit($|\rho_{01}(t)|$) initially in the state$|+\rangle$.Results from 
the 2nd, 4th and 6th order perturbative time-convolution QMEs are 
compared with the numerically exact HEOM results.The parameters 
are set in Table~\ref{tab:bath_params}}
\label{fig:compare_qme}
\end{figure}

We then show that under $1/f$ noise, even high-order time-nonlocal quantum master equations (TNL-QME) fail to capture qubit dephasing dynamics accurately, with errors arising from the truncation inherent to the perturbative approximation.
Based on previous work \cite{chen09a,yan21}, the HEOM
truncated at order $L$th hierarchy is formally equivalent to a time-nonlocal quantum master equation (TNL-QME) at
perturbative order $N=2L$. In Fig.\ref{fig:compare_qme}, we compare the exact HEOM results (which reproduce the analytical solution in Eq.\eqref{eq:pure_dephasing_decoherence}) with the 2nd, 4th, and 6th order TNL-QMEs.

The results indicate that, even for weak system-bath coupling considered in the work, the perturbative TNL-QMEs deviate from the correct dynamics after only a short time. They systematically overestimate the overall coherence time, and the 2nd- and 6th-order expansions also generate spurious
oscillatory components, which leads to the abrupt decay of
$|\rho_{01}(t)|$ and its recovery. Similar effects were
reported in our earlier study of absorption spectra
\cite{chen09a}, where unphysical peaks appeared in the
2nd-order TNL-QME.

This behavior reflects a fundamental limitation of the 
perturbative TNL-QMEs: they are more accurate 
for a rapidly decaying bath
correlation function, while $1/f$ noise corresponds to a
slow bath with long memory. 
We therefore conclude that TNL-QMEs
are generally not reliable for describing qubit decoherence
in the presence of $1/f$ noise, and nonperturbative methods
such as HEOM are required.

\subsection{Dynamical decoupling}
\label{sec:DD}

Beyond single-qubit decoherence, we now investigate the
combined effect of coherent driving and $1/f$ noise
through dynamical decoupling (DD). In particular, we focus
on the Carr–Purcell–Meiboom–Gill (CPMG) 
and Uhrig dynamical decoupling (UDD) sequences, where
multiple $X/Y$ gate pulses are applied to suppress phase
errors from low-frequency noise \cite{viola98,viola99}.

To simulate qubit dynamics under realistic finite-duration
microwave pulses, we first transform the system Hamiltonian into the rotating frame defined by the qubit frequency $\omega_1$ \cite{kwon21}. This is implemented via the unitary operator $R(t)=\exp(i\omega_1\sigma_z t/2)$.  The rotating wave approximation (RWA) then applied, which is well justified under typical superconducting qubit parameters. This yields the following driven Hamiltonian in the rotating frame:
\begin{equation}\label{equ:rot_hamiltonian}
H_S^{\text{rot}} = \frac{\Omega(t)}{2}(\sigma_x\cos{\phi}+\sigma_y\sin{\phi}) \;\;.
\end{equation}
Here, the time-dependent pulse amplitude
$\Omega(t)$ is non-zero only during the application of a pulse. For a $\pi$-rotation of duration $\tau$, we set $\Omega(t)=\pi/\tau$. The phase $\phi$ specifies the rotation axis in the equatorial plane of the Bloch sphere: $\phi=0$ corresponds to an X-gate, while $\phi=\pi/2$ corresponds to a Y-gate. During the waiting intervals between pulses, the drive is switched off, i.e., $\Omega(t)=0$.

Decoherence control with the CPMG sequence is then
investigated. Unless stated otherwise, all simulations begin
with the qubit initialized at $t=0$ in the state
$\rho(0)=|+\rangle\langle +|$. Dephasing is quantified by
tracking the magnitude of the off-diagonal element
$|\rho_{01}(t)|$. Each simulation applies a sequence of 20
X/Y pulses. To ensure the pulses themselves perform as intended, the pulse parameters were first calibrated in the absence of noise. By solving the Schrödinger equation for the driven system(Equ.~\eqref{equ:rot_hamiltonian}), a pulse duration of $\tau=15~\mathrm{ns}$ was determined to implement a high-fidelity $\pi$-rotation around the specified axis (X or Y). The interpulse spacing was set to $\Delta t=118~\mathrm{ns}$. The first
pulse is applied at $t_1=\Delta t/2=59~\mathrm{ns}$ to ensure symmetric refocusing intervals.

\begin{figure}[H]
    \centering
    \includegraphics[width=\linewidth]{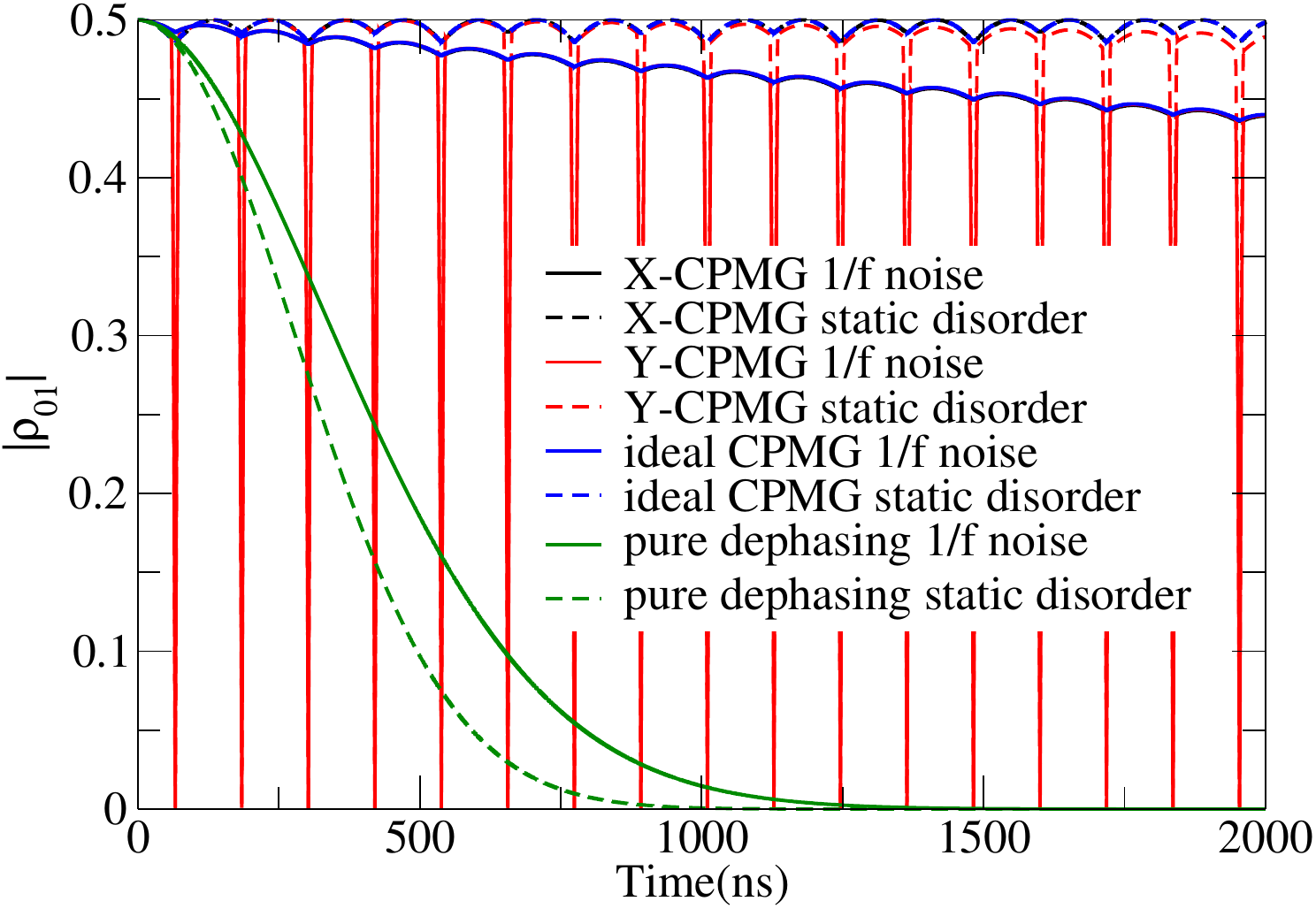}
    \caption{Dynamics of the coherence of a single qubit, $|\rho_{01}|$, under a 20-pulse CPMG sequence. The figure compares the performance of sequences using finite-duration X-gates (X-CPMG, black curves), Y-gates (Y-CPMG, red curves), and idealized instantaneous pulses (ideal CPMG, blue curves). For each sequence, the dynamics under both realistic 1/f noise (solid lines) and an equivalent pure static disorder model (dashed lines) are shown. The free induction decay (pure dephasing, green curves) is included for baseline comparison.}
    \label{fig:CPMG}
\end{figure}

\begin{figure}[H]
    \centering
    \includegraphics[width=\linewidth]{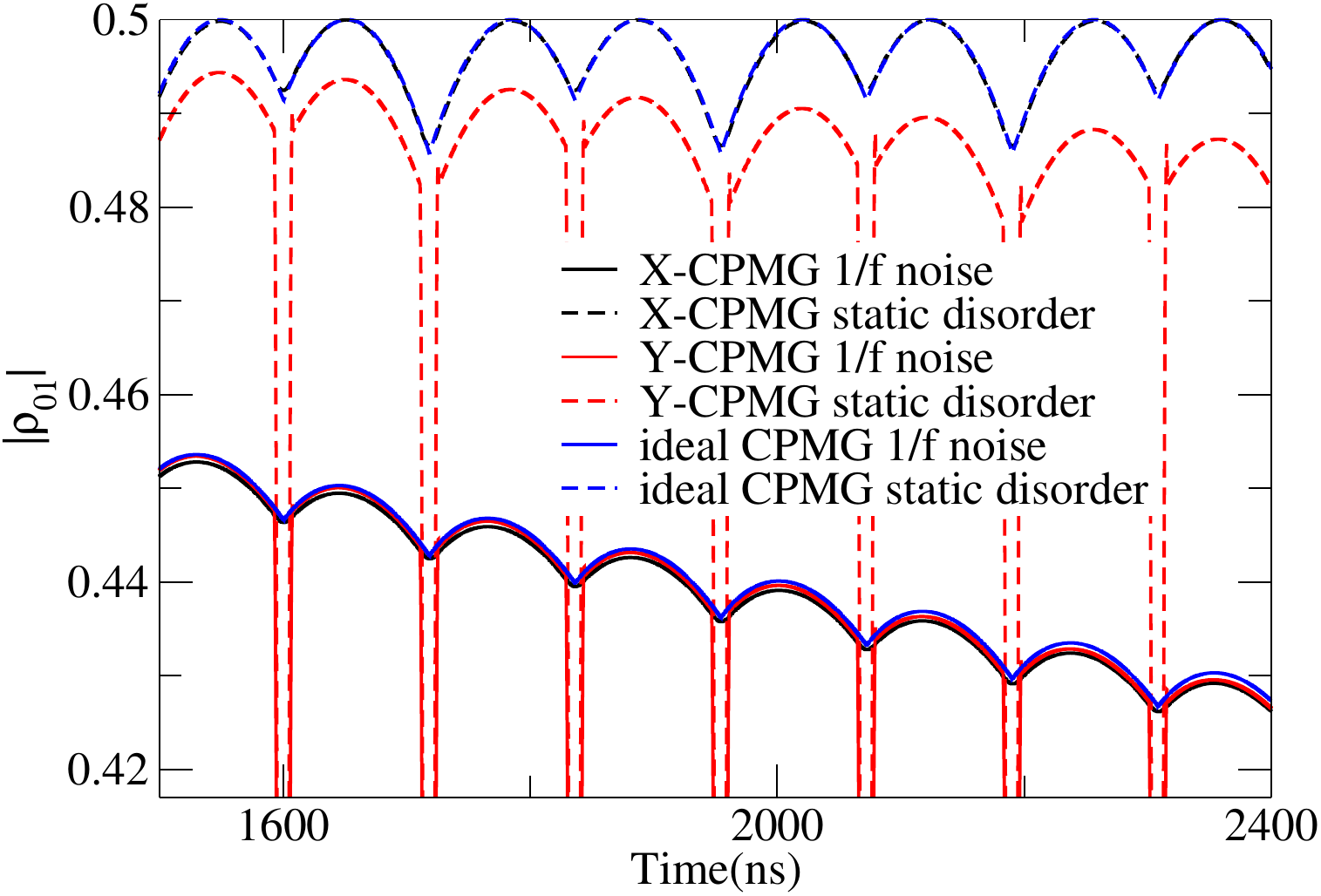}
    \caption{A magnified view of the long-time coherence dynamics from Fig.~\ref{fig:CPMG}, focusing on the decay of the echo peaks.}
    \label{fig:CPMG_2}
\end{figure}

We simulate the dynamics for different sequence types (X-CPMG, Y-CPMG, ideal) under both $1/f$ noise (solid lines) and an equivalent pure static disorder model (dashed lines). The so-called ideal CPMG sequence assumes instantaneous $\pi$-pulses and neglects errors that arise during the finite duration of real pulses. It therefore serves as a benchmark for the best possible performance. As shown in Figs.~\ref{fig:CPMG} and \ref{fig:CPMG_2}, all DD sequences suppress decoherence compared to free induction decay.

However, a more detailed comparison reveals important differences. We first compare the results from the $1/f$ noise model with those from the static disorder model, in the ideal case with impulsive $\pi$-pulses. As shown in Figs.~\ref{fig:CPMG} and \ref{fig:CPMG_2} (blue dashed line), for the static disorder model, errors are strongly suppressed, and the echo amplitude nearly returns to its ideal value at the peak. In contrast, under $1/f$ noise (blue solid line), the error continues to grow with the number of pulses, highlighting the difference between low-frequency $1/f$ noise and static disorder.

When considering finite pulse duration under the static disorder model, X-CPMG sequences show nearly identical behavior to the ideal sequence (overlapping black and blue dashed curves). In contrast, Y-CPMG sequences display a pronounced deviation: the finite-duration result (red dashed line) yields visible lower coherence than the ideal sequence (blue dashed line). This indicates that X- and Y-CPMG sequences yield different results when considering the effect of finite pulse duration.

\begin{figure}[H]
    \centering
    \includegraphics[width=\linewidth]{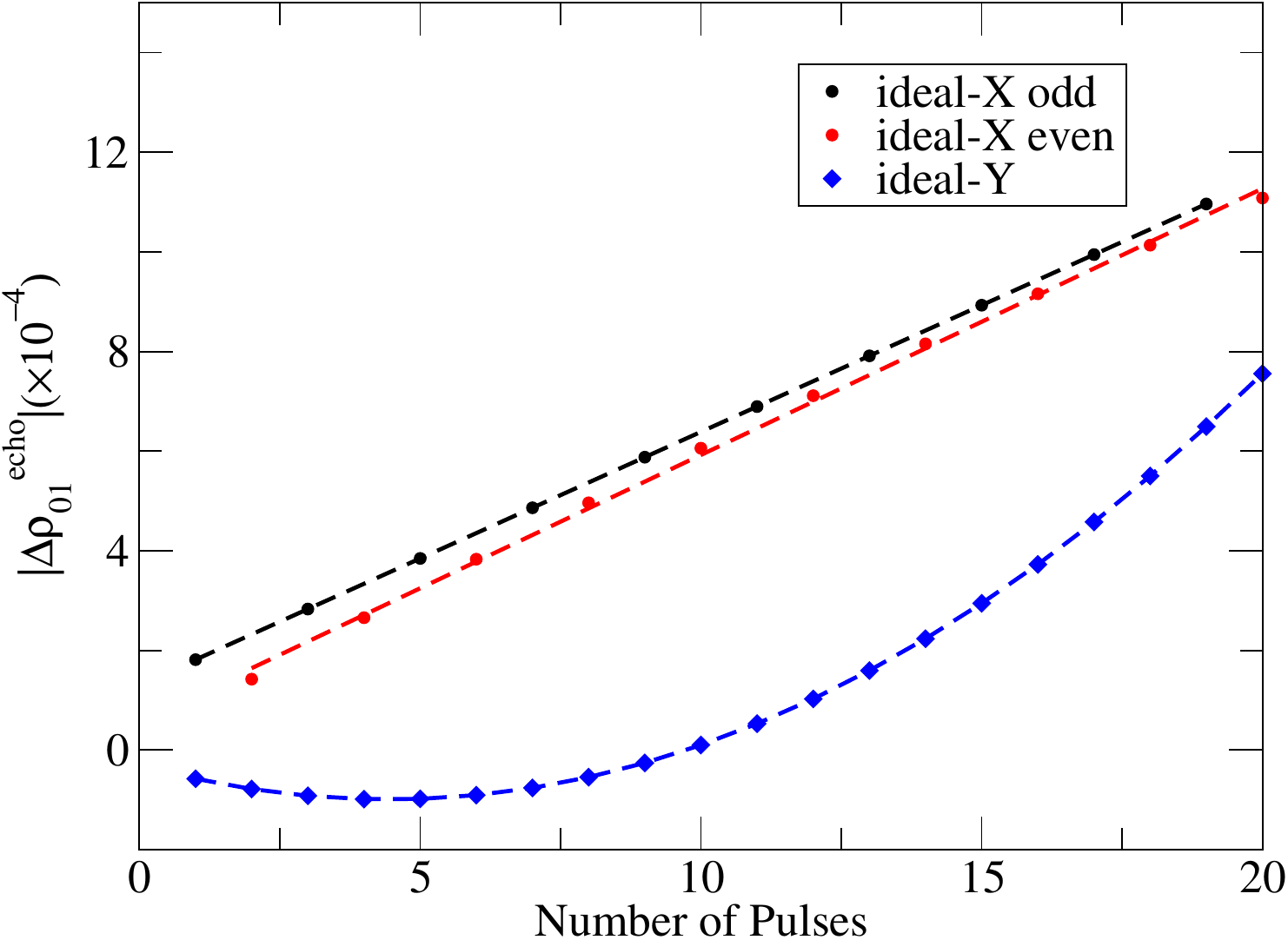}
    \caption{Maximum echo difference $\Delta|\rho_{01}^{\text{echo}}|$ vs. pulse number $n$ for the CPMG sequence. Black circles represent X-CPMG odd pulses, red circles represent X-CPMG even pulses, and blue diamonds represent Y-CPMG. The solid lines are fits to the data, revealing a linear error accumulation for X-CPMG and a quadratic accumulation for Y-CPMG. The fitting functions are:
    \textbf{X-CPMG odd:} $\Delta|\rho_{01}^{\text{echo}}| = (5.081 \times 10^{-5})n + 1.306 \times 10^{-4}$ ($R^2 = 0.995$).
    \textbf{X-CPMG even:} $\Delta|\rho_{01}^{\text{echo}}| = (5.348 \times 10^{-5})n + 5.741 \times 10^{-4}$ ($R^2 = 0.998$).
    \textbf{Y-CPMG:} $\Delta|\rho_{01}^{\text{echo}}| = (3.528 \times 10^{-6})n^2 - (3.131 \times 10^{-5})n - 2.962 \times 10^{-5}$ ($R^2 = 0.994$).}
    \label{fig:ideal}
\end{figure}

We then focus on the effect of finite pulse duration under $1/f$ noise. To characterize the error patterns of different sequences, we evaluate the maximum echo amplitude of the off-diagonal element $|\rho_{01}^{\text{echo}}|$ after each pulse and define the error relative to the ideal sequence as
$\Delta|\rho_{01}^{\text{echo}}| =
|\rho_{01}^{\text{echo, ideal}}| - |\rho_{01}^{\text{echo,
finite}}|$. As shown in Fig.~\ref{fig:ideal}, for X-CPMG under $1/f$ noise, the error increases approximately linearly with the pulse number $n$. In contrast, Y-CPMG follows a quadratic scaling with $n$, indicating that errors accumulate much faster in Y-CPMG sequences at long times.

To better understand the different scaling of $\Delta|\rho_{01}^{\text{echo}}|$ for X- and Y-CPMG sequences under $1/f$ noise, we again compare the results with the static disorder case. As shown in Fig.~\ref{fig:CPMG_2}, under the static disorder model, the error in the X-CPMG sequence is very small, while the Y-CPMG sequence exhibits quadratic error accumulation (red vs. blue dashed lines) as in the case of the $1/f$ noise. This provides strong evidence that the linear error in X-CPMG is caused by the “dynamic” part of the $1/f$ spectrum (slow fluctuations that cannot be treated as static on the sequence timescale), whereas the quadratic error in Y-CPMG is dominated by the “static”, low frequency components of the noise.

It has been found previously that, a small error in the pulse rotation axis (e.g., a small Y-component in an intended X-gate) may lead to an error that accumulates linearly with pulse number for an X-CPMG sequence\cite{xiao11,su12}. The observed linear dependence in X-CPMG is therefore likely a result of such rotation errors combined with the dynamic component of $1/f$ noise.

\begin{figure}[H]
    \centering
    \includegraphics[width=\linewidth]{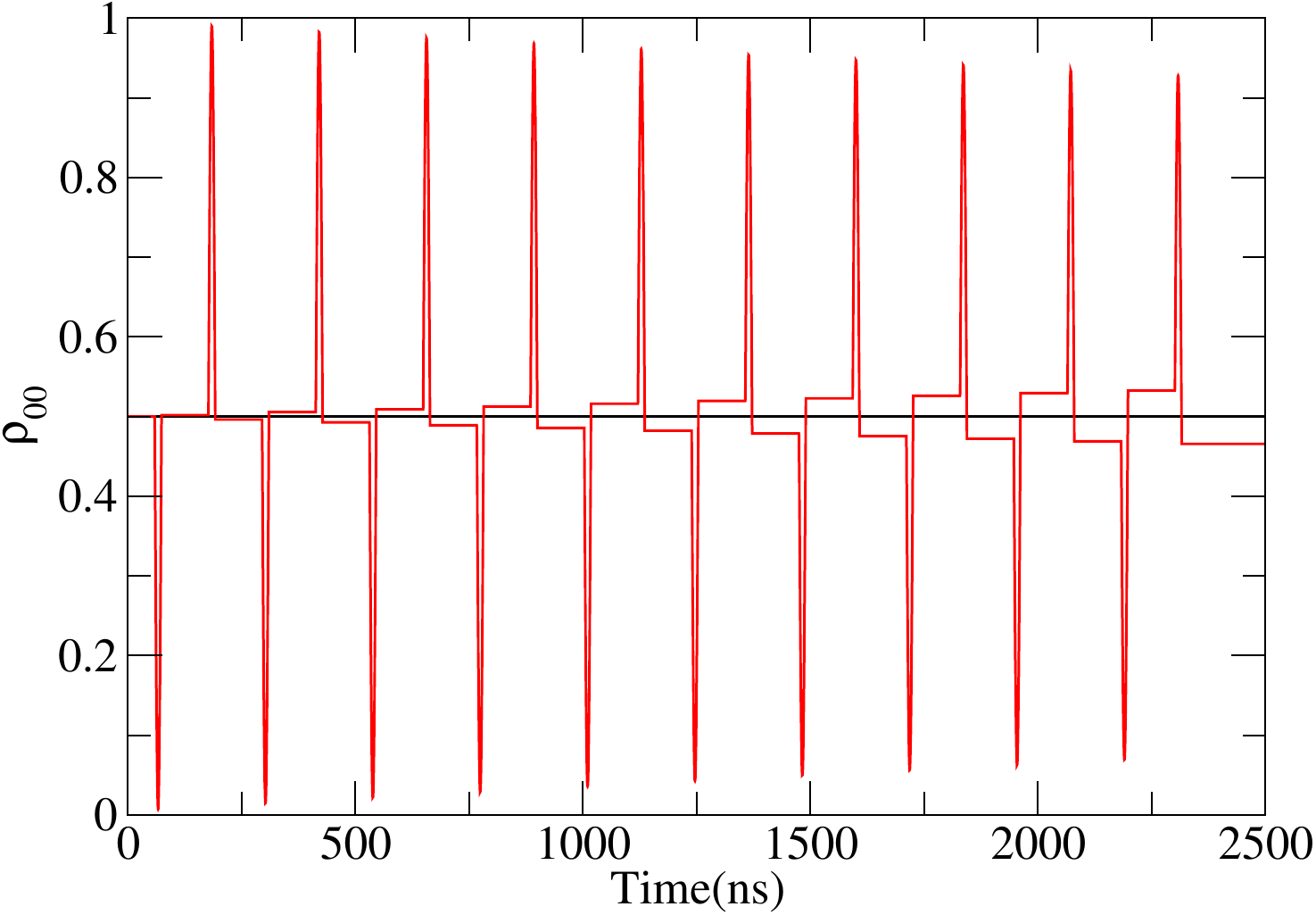}
    \caption{Population dynamics for X-CPMG (black) and Y-CPMG (red).}
    \label{fig:pop}
\end{figure}

For Y-CPMG sequences, we find that they introduce population errors not present in the X-CPMG case. As shown in Fig.~\ref{fig:pop}, the quantum state fails to return exactly to the Bloch-sphere poles, and the ground-state population $\rho_{00}(t)$ deviates from the ideal value of 0.5 after each pulse. This deviation grows approximately linearly with the pulse number $n$, and should be related to the quadratic scaling of $\Delta|\rho_{01}^{\text{echo}}|$ with $n$.
It also explains the small spikes in $|\rho_{01}(t)|$ observed for Y-CPMG in Fig.~\ref{fig:CPMG_2}, corresponding to times when the $|0\rangle$ and $|1\rangle$ populations become equal.

\begin{figure}[H]
    \centering
    \includegraphics[width=\linewidth]{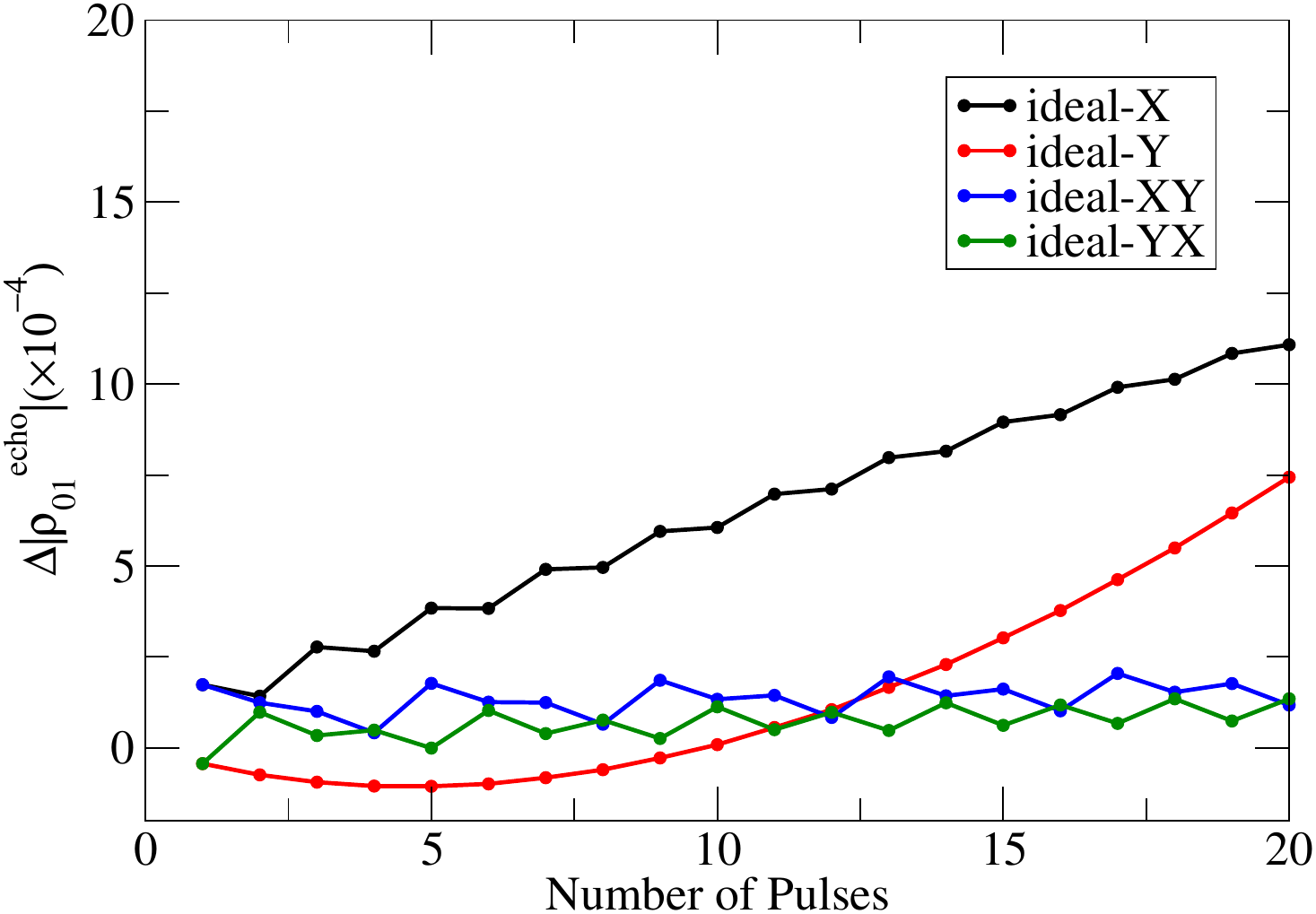}
    \caption{Maximum echo for XY/YX-CPMG vs. pulse number $n$. Black and red curves reproduce X/Y-CPMG data from Fig.~\ref{fig:ideal} for comparison.}
    \label{fig:XY_pulse}
\end{figure}

\begin{figure}[H]
    \centering
    \includegraphics[width=\linewidth]{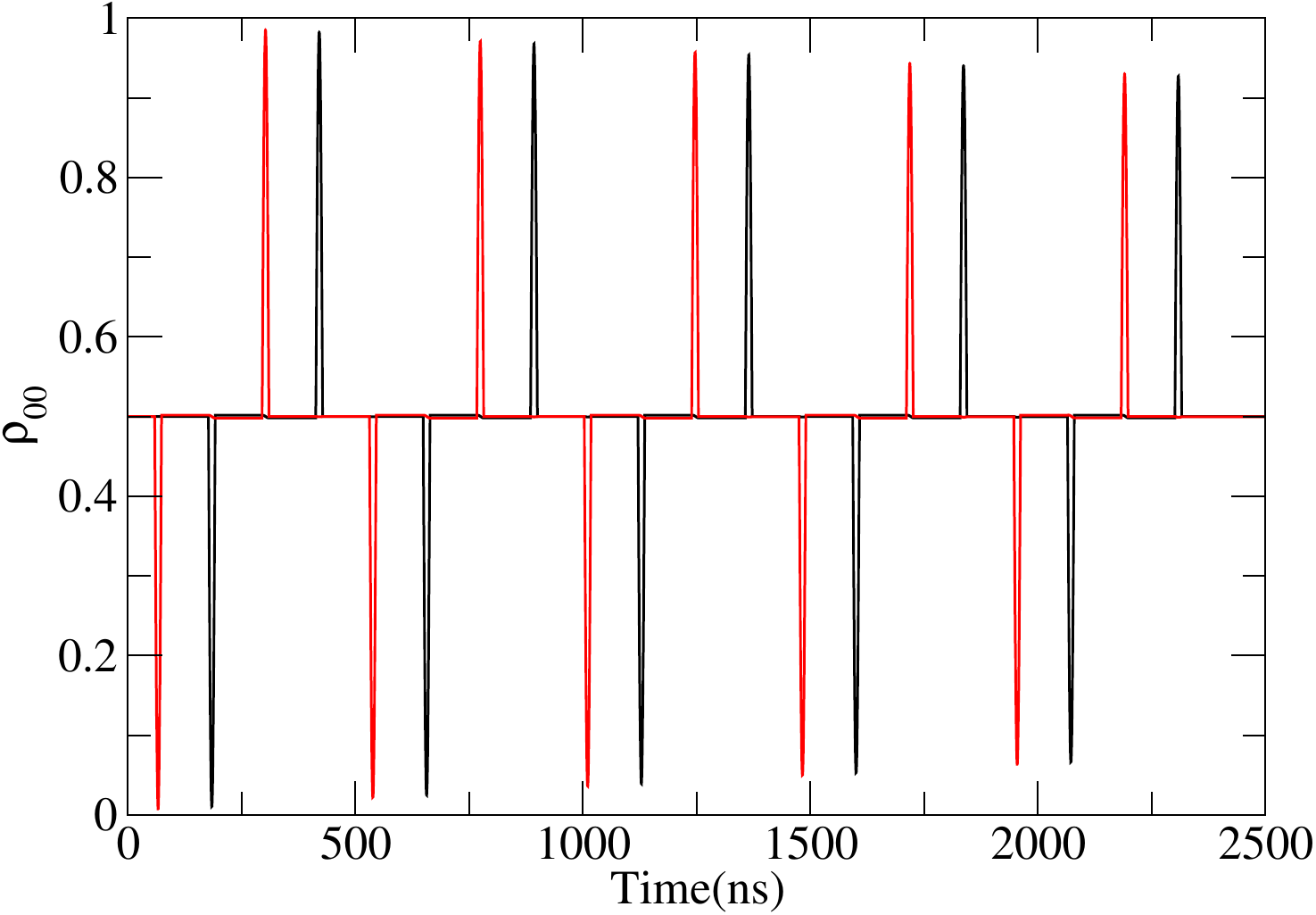}
    \caption{Population dynamics for XY-CPMG (black) and YX-CPMG (red).}
    \label{fig:XY_pulse_population}
\end{figure}

Beyond the simple X- and Y-CPMG sequences, we also consider multi-axis sequences where X and Y pulses alternate (denoted XY-CPMG and YX-CPMG). As shown in Fig.~\ref{fig:XY_pulse}, the corresponding error $\Delta|\rho_{01}^{\text{echo}}|$ is strongly suppressed at long times. Fig.~\ref{fig:XY_pulse_population} further confirms that XY- and YX-CPMG also significantly reduce population errors, although the quantum state still does not pass exactly through the Bloch-sphere poles. Overall, alternating-axis sequences achieve superior performance by compensating for both coherence and population errors.
 
\begin{figure}[H]
    \centering
    \includegraphics[width=\linewidth]{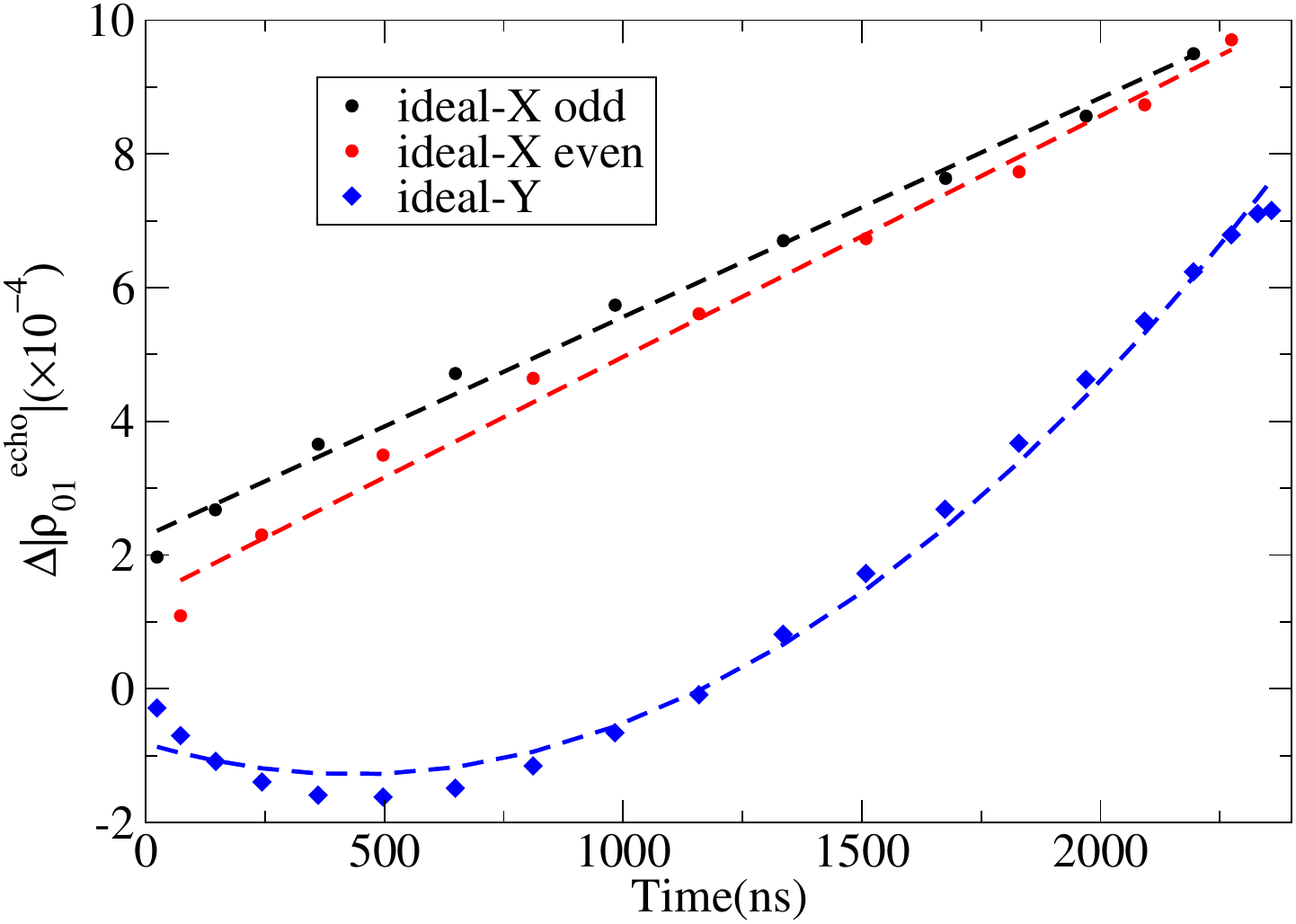}
    \caption{Maximum echo difference $\Delta|\rho_{01}^{\text{echo}}|$ vs. time $t$ for the UDD sequence. Symbols follow the convention of Fig.~\ref{fig:ideal}. The solid lines are fits to the data, confirming that the scaling laws observed in CPMG are general. The fitting functions are:
    \textbf{X-UDD odd:} $\Delta|\rho_{01}^{\text{echo}}| = (3.278 \times 10^{-7})t + 2.283 \times 10^{-4}$ ($R^2 = 0.992$).
    \textbf{X-UDD even:} $\Delta|\rho_{01}^{\text{echo}}| = (3.606 \times 10^{-7})t + 1.358 \times 10^{-4}$ ($R^2 = 0.991$).
    \textbf{Y-UDD:} $\Delta|\rho_{01}^{\text{echo}}| = (2.411 \times 10^{-10})t^2 - (2.111 \times 10^{-7})t - 8.198 \times 10^{-5}$ ($R^2 = 0.993$).}
    \label{fig:UDD_ideal}
\end{figure}

Finally, we examine Uhrig Dynamical Decoupling (UDD)
\cite{Uhrig07}, which employs non-uniform pulse intervals.
When plotting $\Delta|\rho_{01}^{\text{echo}}|$ against the total time $t$ rather than the pulse number $n$ (in Fig.~\ref{fig:UDD_ideal}), we find that UDD and CPMG follow the same scaling laws: the error accumulates linearly with time for X-UDD and quadratically for Y-UDD.
These results indicate that such scaling is general for both uniform and non-uniform sequences. 


\subsection{Operation the CR Gate under 1/f noise}
\label{sec:CR}

In this subsection, we extend the HEOM framework to study a two-qubit gate under $1/f$ noise, using the cross-resonance (CR) gate as a representative example. High-fidelity two-qubit operations are essential for scalable quantum computing architectures \cite{arute19,chow14}. The CR gate is implemented by applying a microwave drive to the control qubit at the transition frequency of the target qubit, thereby inducing an effective ZX interaction mediated by their fixed capacitive coupling \cite{megesan20,ware19,tripathi19,malekakhlagh20}. Because it requires only fixed-frequency qubits, fixed coupling, and standard microwave pulses, the CR gate is also realteively simple to implement \cite{chow11,paraoanu06,rigetti10}.

To analyze the gate dynamics, we transform the lab-frame Hamiltonian of two capacitively coupled qubits into a doubly rotating frame \cite{rigetti10}. This frame co-rotates with the drive frequency $\omega_d$ applied to the control qubit (qubit 1) and with the transition frequency of the target qubit (qubit 2). For the CR gate, we set $\omega_d = \omega_2$. After also applying the rotating wave approximation (RWA) to remove fast-oscillating terms, the system Hamiltonian $H_S$ takes the form:
\begin{equation}\label{equ:CR}
    H_{\text{CR}}^{\text{rot}} = \frac{\Delta}{2}\sigma_z^{(1)} + g(\sigma_x^{(1)}\sigma_x^{(2)} + \sigma_y^{(1)}\sigma_y^{(2)}) + \frac{\Omega(t)}{2}\sigma_x^{(1)}.
\end{equation}
Here, $\Delta = \omega_1 - \omega_2$ is the detuning between the control and target qubit frequencies,
$g$ is their static XY coupling strength, and $\Omega(t)$ is the microwave drive amplitude on the control qubit.
We assume each qubit is independently coupled to its own local bath causing $Z$-type noise, as described in the previous sections.

In the dispersive regime ($|\Delta| \gg g, \Omega(t)$), the Schrieffer–Wolff (S–W) transformation can be applied to derive an effective Hamiltonian \cite{megesan20}. This transformation shows that the leading interaction in the CR gate is of the ZX type, with an effective coupling rate $\Omega_{ZX} \approx \tfrac{g\Omega(t)}{\Delta}$. The entangling rotation angle is therefore determined by the time integral of this rate. To implement a target rotation angle $\theta$, the drive pulse must satisfy
\begin{equation}\label{equ:amplitude_cr}
\int_0^\tau \frac{g \Omega(t)}{\Delta} dt = \theta,
\end{equation}
where $\tau$ is the pulse duration. In practice, this ideal condition is typically not met exactly. Nevertheless, Eq.~\eqref{equ:amplitude_cr} provides a useful guideline and a reliable starting point for calibrating realistic CR gates in simulations and experiments.

A set of optimized parameters For the two-qubit CR gate is first obtained by maximizing gate fidelity in the absolute of environmental noise.
For this purpose, we employ an iterative parameter-screening procedure. First, we conduct a coarse search over detuning $\Delta$, drive amplitude $\Omega$, pulse duration $\tau$, and the axes/angles of single-qubit calibration gates. At this stage, 
the choice of $\tau$ and $\Omega$ is constrained by Eq.~\eqref{equ:amplitude_cr}. The resulting calibrated unitary, $U_{\text{calibration}}$, is then compared with the target gate the ideal target unitary,
$U_{\text{ideal}}=\exp{\left(-\frac{i\pi}{4}\sigma_z^{(1)}\sigma_x^{(2)}\right)}$,
by projecting their difference onto the Pauli basis.

If the dominant discrepancies correspond to single-qubit terms (e.g., IZ, ZI), we correct them by adding appropriate single-qubit gates before and after the CR evolution governed by $H_{\text{CR}}^{\text{rot}}$.
If they correspond to two-qubit entangling terms (e.g., ZX, XY), we release the constraint on Eq.~\eqref{equ:amplitude_cr} and perform a finer search over $\Delta$, $\tau$, and $\Omega$. This loop is repeated until the gate fidelity exceeds 0.999 or a maximum number of iterations is reached.

The optimized parameters are listed in Table~\ref{tab:system_params_CR}. With the inter-qubit coupling fixed at $g/2\pi=50~\mathrm{MHz}$, the optimization yields $\Delta/2\pi = 0.5148~\mathrm{GHz}$, $\tau = 132~\mathrm{ns}$, and $\Omega/2\pi = 105.63~\mathrm{MHz}$. Calibration is completed with corrective single-qubit $R_z$ rotations: for the control qubit (qubit 1), $U_{\text{pre}}=-0.750050\pi$ and $U_{\text{post}}=-0.093750\pi$; for the target qubit (qubit 2), the same rotation angle $0.593800\pi$ is applied both before and after the CR interaction. The total calibrated unitary evolution operator is thus given by the sequence $U_{\text{calibrated}}(t) = U_{\text{post}} U_{\text{CR}}(t) U_{\text{pre}}$, where $U_{\text{pre/post}}$ are the products of the single-qubit corrective rotations and $U_{\text{CR}}(t)$ is the propagator generated by $H_{\text{CR}}^{\text{rot}}$.
({\color{red} \bf ??
1. are Upred and Upost instantaneous? 2. Do we need to add a figure? 
 ??}) 



We then characterize the performance of the CR gate in the presence of $1/f$ noise. For this purpose, we first construct the Choi matrix $\chi_\mathcal{E}$ corresponding to the quantum channel $\mathcal{E}$ that fully describes the noisy gate: 
\begin{equation}
\chi_\mathcal{E} = \sum_{i,j} |i\rangle\langle j| \otimes \mathcal{E}(|i\rangle\langle j|),
\end{equation}
where ${|i\rangle}$ denotes a basis for the two-qubit Hilbert space.

Strictly speaking, only the value at the final time corresponds to the gate fidelity. Here, the noisy Choi matrix $\chi_{\text{noisy}}(t)$ is compared against the noiseless, calibrated evolution $\chi_{\text{calibrated}}(t)$, whose propagator is given by $U_{\text{calibrated}}(t) = U_{\text{post}} U_{\text{CR}}(t) U_{\text{pre}}$. This dynamic comparison effectively demonstrates the accumulation of errors on the CR gate due to noise.

We evaluate the time-resolved gate fidelity using the Choi representation,
\begin{equation}  
\mathcal{F}_{\text{gate}}(t) = \frac{\text{Tr}(\chi_{\text{noisy}}(t) \chi_{\text{calibrated}}(t))}{\text{Tr}(\chi_{\text{calibrated}}^2(t))}.
\end{equation}
Here, the noisy channel $\chi_{\text{noisy}}(t)$ is compared against the noiseless, calibrated evolution $\chi_{\text{calibrated}}(t)$.
Strictly, only the value at the final gate time corresponds to the gate fidelity; the intermediate $\mathcal{F}_{\text{gate}}(t)$ serves as a diagnostic of error accumulation under the noisy environment.

\begin{figure}[H]
    \centering
    \includegraphics[width=\linewidth]{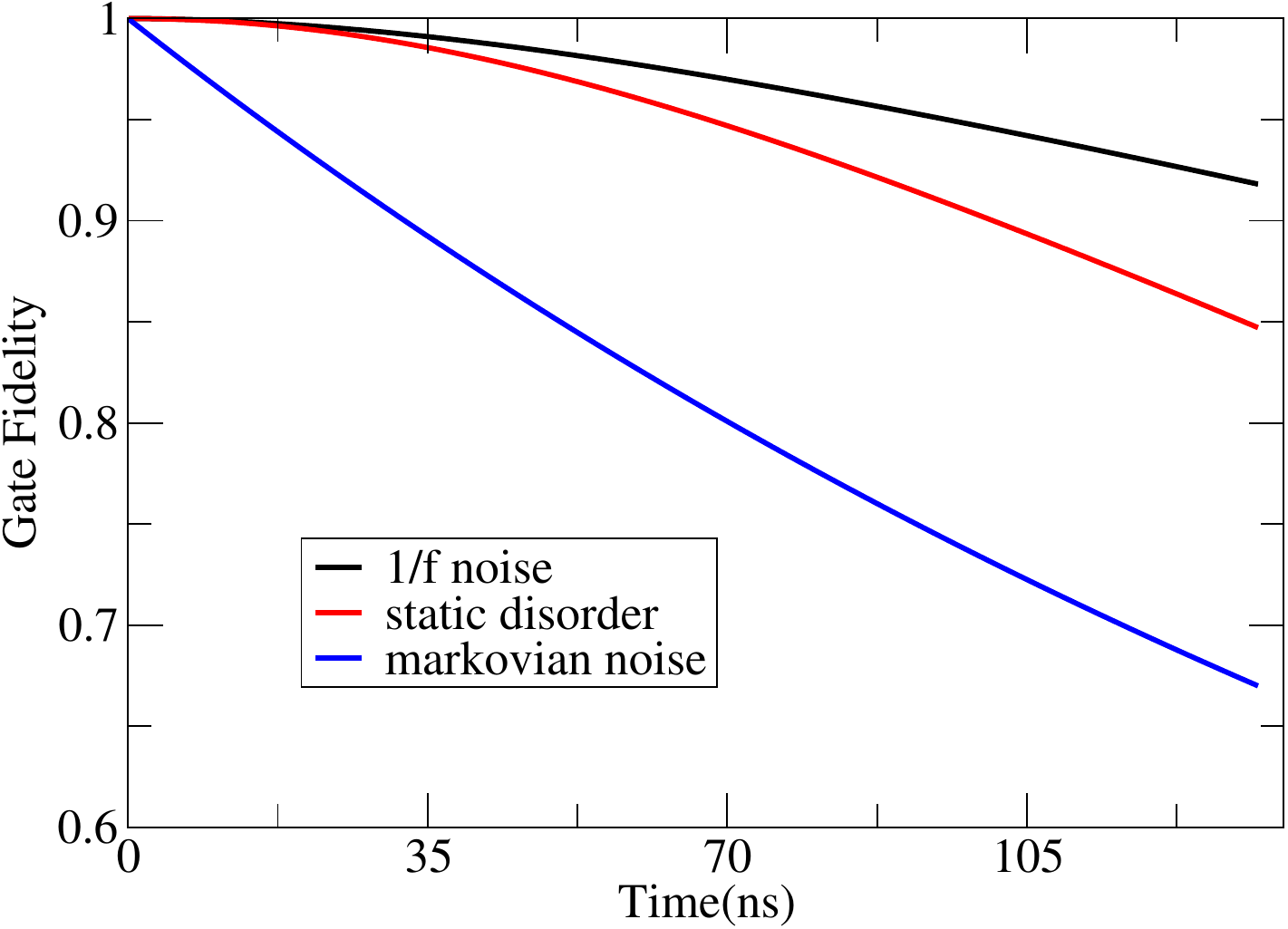}
    \caption{Gate fidelity dynamics under different Z-type noise.}
    \label{fig:CR_gate_fidelity}
\end{figure}

As shown in Fig.~\ref{fig:CR_gate_fidelity}, we compare three noise models: $1/f$ noise, static disorder, and Markovian dephasing. For a fair comparison, the integrated noise power $\int_0^\infty S(\omega)\,d\omega$ is matched between the $1/f$ and static-disorder cases, while the Markovian case uses a Lindblad model with pure-dephasing time $T_\phi \approx 576~\text{ns}$ chosen to yield comparable overall dephasing strength. The resulting dynamics are qualitatively distinct: $1/f$ noise and static disorder produce an initial quadratic decay in fidelity, characteristic of environments with long correlation times, whereas the Markovian model exhibits an exponential decay. This contrast highlights that the non-Markovian nature of $1/f$ noise cannot be captured by a simple Markovian approximation.

For a more detailed assessment of gate performance, we project the final Choi matrix onto the Pauli Transfer Matrix (PTM) basis, following Refs.~\cite{chow12,mckay16}. The PTM, denoted $\mathcal{R}$, fully characterizes the quantum process in the Pauli basis, with matrix elements defined as
\begin{equation}
\mathcal{R}_{ij} = \frac{1}{4} \text{Tr}\left[ P_i \mathcal{E}(P_j) \right],
\end{equation}
where $P_i, P_j$ are the two-qubit Pauli operators.

\begin{figure}[H]
    \centering
\includegraphics[width=\linewidth]{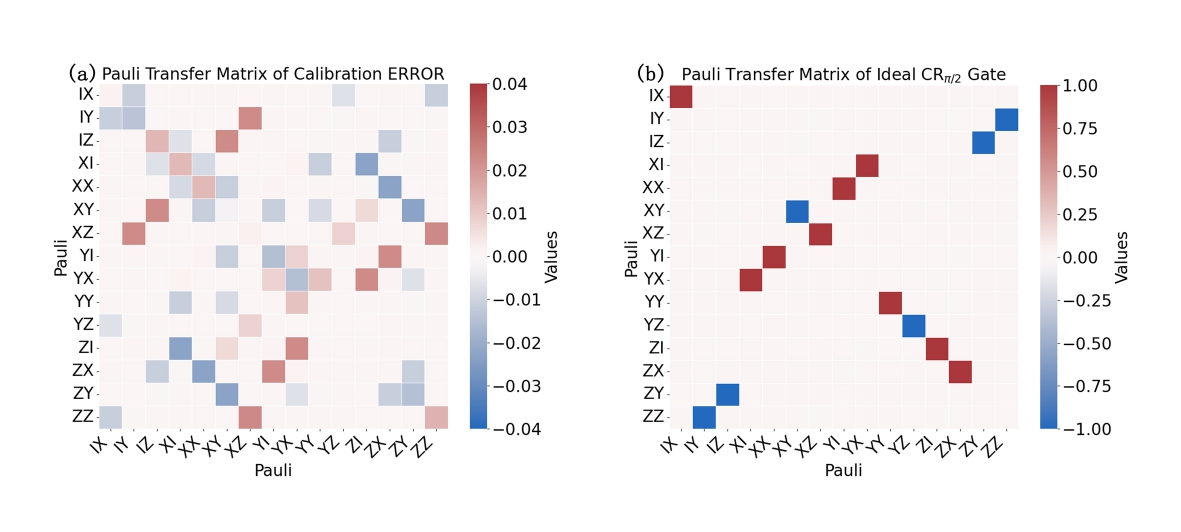}
    \caption{Pauli Transfer Matrix (PTM) for the ideal and calibrated noiseless $\text{CR}_{\pi/2}$ gate. 
    \textbf{(a)} The coherent error PTM of the calibrated gate, defined as $\Delta\mathcal{R} = \mathcal{R}_{\text{calibrated}} - \mathcal{R}_{\text{ideal}}$. It quantifies the small, residual errors inherent to the gate's physical implementation. 
    \textbf{(b)} The PTM for the ideal $\text{CR}_{\pi/2}$ gate. }
    \label{fig:PTM_ideal}
\end{figure}

Fig.~\ref{fig:PTM_ideal}(b) shows the PTM for the ideal $\text{CR}_{\pi/2}$ gate, whose structure highlights the entangling nature of the operation. Especially, the off-diagonal elements show the intended transformations of Pauli operators, such as the mapping of IY to -ZY.
To analyze deviations from the ideal gate, we separate the total error into two contributions. Fig.~\ref{fig:PTM_ideal}(a) shows the intrinsic \textit{coherent error} of the gate, defined as $\Delta\mathcal{R} = \mathcal{R}_{\text{calibrated}} - \mathcal{R}_{\text{ideal}}$. This term captures the small residual deviation of the best calibrated, noiseless gate from the ideal gate, originating from unwanted interactions inherent to the CR Hamiltonian.

\begin{figure}[H]
    \centering
    \includegraphics[width=\linewidth]{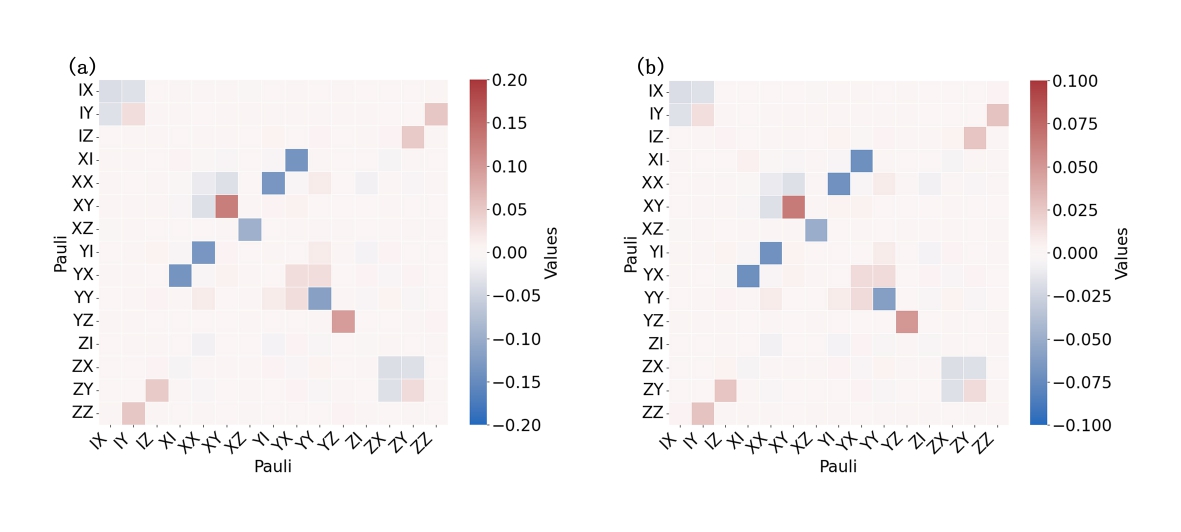}
 \caption{Comparison of net error PTMs for the $\text{CR}_{\pi/2}$ gate (defined as $\Delta\mathcal{R} = \mathcal{R}_{\text{noise}} - \mathcal{R}_{\text{calibrated}}$) induced by different noise models. 
    \textbf{(a)} The net error PTM for a total static disorder model with the same integrated noise power. 
    \textbf{(b)} The net error PTM from the full 1/f noise model . A direct comparison shows that while the static disorder model in (a) captures the general error patterns, the full 1/f noise in (b) induces errors of a significantly larger magnitude. This highlights the crucial impact of the dynamic components of the noise, which lead to stronger decoherence (more negative diagonal elements) and more pronounced coherent errors (off-diagonal elements).}
    \label{fig:PTM_diff2}
\end{figure}

To isolate the impact of the dissipative environment, we compute the net error PTM induced by the 1/f noise, $\Delta\mathcal{R} = \mathcal{R}_{\text{noise}} - \mathcal{R}_{\text{calibrated}}$, shown in Fig.~\ref{fig:PTM_diff2}(b). It can be seen that a few diagonal and off-diagonal elements are significantly affected by the $1/f$ noise. The negative values along the diagonal, particularly for terms like IX, IY, and XY, represent the decay of these Pauli components due to dephasing. The non-zero element at (XY, YY) indicates that the noise induces an unwanted, erroneous rotation that maps the YY state partially onto the XY state. 

We also calculate the error PTM under the static disorder model, shown in Fig.~\ref{fig:PTM_diff2}(a). The overall pattern is very similar to the $1/f$ noise case, but the magnitudes of the errors are generally larger. This indicated that the error pattern of the $1/f$ noise is actually very close to that of an equivalent static disorder model for the CR gate.



\section{Conclusions and discussions}
\label{sec:conclusion}

In this work, we extend the HEOM framework to simulate the dynamics of realistic single- and two-qubit gates in superconducting hardware under the influence of non-Markovian $1/f$ noise. Our key findings can be summarized as follows: 
First, we investigated the parameters for modeling $1/f$ noise and demonstrated the critical role of the low-frequency cutoff, $\omega_{lc}$, in accurately capturing the qubit's decoherence dynamics.

In studying single-qubit dephasing dynamics, we first validate the HEOM framework for treating $1/f$ noise by discussing the choice of a critical low-frequency cutoff parameter. We then show that the commonly used second-order TNL-QME fails to accurately reproduce qubit coherence decay under $1/f$ noise, and may 
introduce spurious oscillating frequency. This is an 
inherent problem of the perturbative TNL-QMEs in the presence of slow baths, and simply increasing the perturbative order does not resolve the issue.


We also use dynamical decoupling to demonstrate the interplay between external pulse driving and the non-Markovian 1/f noise. For both CPMG and UDD sequences,
finite pulse duration leads to additional error accumulation
when using single-axis sequences (all-X or all-Y). 
X-gate sequences exhibit an error that grows approximately linearly with the pulse number, while Y-gate sequences show a quadratic growth. We attribute this behavior to different underlying mechanisms: in X-CPMG sequences, errors are dominated by the dynamic component of the $1/f$ noise, while in Y-CPMG sequences they arise primarily from the static component.

This observation is supported by the control simulation using a \textit{total static disorder} model with equivalent integrated noise power. In this case, the linear error of the X-CPMG sequence was completely suppressed, while the quadratic error of the Y-CPMG sequence persists.
We also confirm that alternating the X and Y pulses (e.g., XY- or YX-CPMG) significantly suppresses the overall error accumulation, by being able to compensate for both coherence and population errors. This finding is consistent with the experimental observation\cite{bylander11} and highlights the practical advantage of multi-axis control.



The HEOM framework was also extended to simulate the more challenging two-qubit CR gate. We constructed the full Choi matrix for the noisy gate and utilized the PTM formalism to present detailed error patterns of 1/f noise. The PTM analysis indicate
that incoherent errors are primarily localized in a few diagonal and off-diagonal elements of
the PTM, which is very similar to the error pattern of an equivalent static disorder model. This finding may be useful for developing methods to improve the performance of two-qubit gates under $1/f$ noise.


In summary, we have demonstrated that the HEOM method provides a powerful and reliable approach for simulating superconducting qubits and quantum gates in the presence of $1/f$ noise.
Besides clarifing some key aspects of $1/f$ noise modeling, out study also provides a robust and detailed simulation protocol for diagnosing errors in both single- and multi-qubit operations.
In future works, the approach presented here can be readily extended to study qubits embedded in more complex environments, providing valuable insights for the design of noise-resilient quantum gates and quantum error-correction strategies.

\begin{acknowledgments}
This work is supported by NSFC (Grant No. 22433006).
We acknowledge helpful discussions with 
Dr. Yujia Zhang, Prof. Ruixia Wang, and Prof. Fei Yan from BAQIS.
\end{acknowledgments}

\bibliography{../main.bib}

\pagebreak
\begin{table}[H]
    \centering
    \caption{Bath parameters for $1/f$ noise.}
    \label{tab:bath_params}
    \begin{tabular}{lc}
        \hline
        Parameter & Value \\
        \hline
        Coupling strength $\eta/2\pi$ & $10^{-7}$ \\
        Noise exponent $s$ & 0 \\
        High-frequency cutoff $\omega_{hc}/2\pi$ & 10 GHz \\
        low-frequency cutoff  $\omega_{hc}/2\pi$ & 0.1 MHz\\
        Temperature $T$ & 50 mK \\
        \hline
    \end{tabular}
\end{table}

\begin{table}[H]
    \centering
    \caption{System-bath coupling parameters under different
    low-frequency cutoffs and PSD definitions.
    $\eta_{\text{c}}$: \textit{cutoff only} schemes;
    $\eta_{\text{csd}}$: \textit{cutoff plus static disorder} schemes; $\sigma^2$:
    Variance of static disorder in \textit{cutoff plus static disorder} schemes.}
    \label{tab:cutoff_params}
    \begin{tabular}{l|cccc}
        \hline
        Parameter & \multicolumn{4}{c}{$\omega_{lc}/2\pi$} \\
        \hline
         & 1Hz & 100Hz & 10 kHz & 1 MHz  \\
        \hline
        $\eta_{\text{c}}$ & $5.328 \times 10^{-8}$ &
        $6.555 \times 10^{-8}$ & $1.000 \times 10^{-7}$&$1.215\times 10^{-7}$ \\
        $\eta_{\text{csd}}$ & $5.120 \times 10^{-8}$ &
        $6.243 \times 10^{-8}$ & $8.515 \times 10^{-8}$ &$ 1.112\times10^{-7}$ \\
        $\sigma^2$ &$5.120 \times 10^{-7}$ & $6.555 \times 10^{-7}$ &$8.515 \times 10^{-7}$ & $ 1.112\times10^{-6}$\\
        $\langle\delta\omega_z^2\rangle$ & $1.311 \times
        10^{-5}$ & $1.311 \times 10^{-5}$ & $1.311 \times 10^{-5}$ & $1.311 \times 10^{-5}$ \\
        
        \hline
    \end{tabular}
\end{table}

\pagebreak
\begin{table}[H]
    \centering
    \caption{Optimized parameters for CPMG pulse sequence of a single qubit system.}
    \label{tab:system_params_CPMG}
    \begin{tabular}{lc}
        \hline
        Parameter & Value \\
        \hline
        Pulse duration $\tau$ & 15 ns \\
        inter-pulse spacing $\Delta t$ & 118 ns\\
        1st pulse start time $t_1$ & 59 ns\\
        The number of CPMG pulses $n_{pulse}$ & 20\\
        \hline
    \end{tabular}
\end{table}

\begin{table}[H]
    \centering
    \caption{Optimized and calibrated system parameters for the CR gate.}
    \label{tab:system_params_CR}
    \begin{tabular}{lc}
        \hline
        \textbf{Parameter} & \textbf{Value} \\
        \hline
        Detuning $\Delta/2\pi$ & 0.5148 GHz \\
        Coupling strength $g/2\pi$ & 50 MHz \\
        Pulse duration $\tau$ & 132 ns \\
        Drive amplitude $\Omega/2\pi$ & 105.6 MHz \\
        \hline
        \multicolumn{2}{c}{\textbf{Single-Qubit R$_z$ Corrections (angles in units of $\pi$)}} \\
        \hline
        Pre-rotation angle (Qubit 1), $\theta_{\text{pre,1}}/\pi$ & -0.750050 \\
        Post-rotation angle (Qubit 1), $\theta_{\text{post,1}}/\pi$ & -0.093750 \\
        Pre-rotation angle (Qubit 2), $\theta_{\text{pre,2}}/\pi$ & 0.593800 \\
        Post-rotation angle (Qubit 2), $\theta_{\text{post,2}}/\pi$ & 0.593800 \\
        \hline
    \end{tabular}
\end{table}

\pagebreak

\pagebreak

\pagebreak

\pagebreak

\pagebreak

\pagebreak

\pagebreak

\pagebreak

\pagebreak

\pagebreak

\pagebreak

\pagebreak

\end{document}